\documentclass[nofootinbib,notitlepage,superscriptaddress,10pt,aps,pra,twocolumn]{revtex4-1}
\usepackage[utf8x]{inputenc}
\usepackage{amsmath}
\usepackage{amssymb}
\usepackage{graphicx}
\usepackage[normalem]{ulem}
\usepackage{epsf,epsfig}
\usepackage{psfrag}
\usepackage{color}
\usepackage{multirow}
\usepackage{diagbox}

\makeatletter
\newcommand\stroke[1]{\mathpalette\stroke@aux{#1}}
\def\stroke@aux#1#2{%
  \ooalign{%
    \hfil$#1^{\;\, \_\hspace{-0.05cm}\_}$\hfil\cr
    \hfil$#1#2$\hfil\cr
  }%
}
\makeatother
\newcommand\dtagliato{\stroke{d}}

\begin{document}

\title{IceCube and GRB neutrinos propagating in quantum spacetime}
%\maketitle

\author{Giovanni Amelino-Camelia}
\affiliation{Dipartimento di Fisica, Universit\`a di Roma ``La Sapienza", P.le A. Moro 2, 00185 Roma, Italy}
\affiliation{INFN, Sez.~Roma1, P.le A. Moro 2, 00185 Roma, Italy}
\author{Leonardo Barcaroli}
\affiliation{Dipartimento di Fisica, Universit\`a di Roma ``La Sapienza", P.le A. Moro 2, 00185 Roma, Italy}
\author{Giacomo D'Amico}
\affiliation{Dipartimento di Fisica, Universit\`a di Roma ``La Sapienza", P.le A. Moro 2, 00185 Roma, Italy}
\affiliation{INFN, Sez.~Roma1, P.le A. Moro 2, 00185 Roma, Italy}
\author{Niccol\'{o} Loret}
\affiliation{Institut Ru\!$\dtagliato$\!er Bo\v{s}kovi\'{c}, Bijeni\v{c}ka cesta 54, 10000 Zagreb, Croatia}
\author{Giacomo Rosati}
\affiliation{Institute for Theoretical Physics, University of Wroc\l{}aw, Pl. Maksa Borna 9, Pl-50-204 Wroc\l{}aw, Poland}

\begin{abstract}
Two recent publications have reported intriguing analyses, tentatively suggesting that some aspects of IceCube
 data might be manifestations of quantum-gravity-modified laws of propagation for neutrinos.
  We here propose a strategy of data analysis which has the advantage of being applicable to several alternative possibilities
    for the laws of propagation of neutrinos in a quantum spacetime. In all scenarios here of interest
    one should  find a correlation between the energy of an observed neutrino and the
    difference between the time of observation of that neutrino and the trigger time of a GRB.
   We select accordingly some GRB-neutrino candidates among IceCube events, and our data analysis finds a rather strong such correlation.
   This sort of studies naturally lends itself to the introduction of a ``false alarm probability", which for our analysis we estimate conservatively to be of 1$\%$. We therefore argue that our findings should motivate a vigorous program of investigation following the strategy here advocated.
\end{abstract}
\maketitle

\section{INTRODUCTION}
The prediction of a neutrino emission associated with gamma ray bursts (GRBs)
is  generic within the  most widely accepted astrophysical models~\cite{fireball}.
After a few years of operation IceCube still reports \cite{icecubeUPDATEgrbnu}
no conclusive detection of GRB  neutrinos,
contradicting some influential predictions \cite{waxbig,meszabig,dafnebig,otherbig} of the GRB-neutrino observation rate by IceCube.
Of course, it may well be the case that the efficiency of neutrino production at GRBs  is much lower
than had been previously estimated~\cite{small1,small2,small3}. However, from the viewpoint of quantum-gravity/quantum-spacetime
research it is interesting to speculate that the IceCube results for GRB neutrinos might be misleading
because of the assumption that GRB neutrinos should be detected in very close temporal coincidence with
the associated $\gamma$-rays:
a sizeable mismatch between GRB-neutrino detection time and trigger time for the GRB is expected
in several much-studied models of neutrino propagation in a quantum spacetime
(see Refs.\cite{gacLRR,jacobpiran,gacsmolin,grbgac,gampul,urrutia,gacmaj,myePRL,gacGuettaPiran,steckerliberati} and references therein).

This possibility was preliminarily explored in Ref.\cite{gacGuettaPiran}
using only IceCube data from April 2008 to May 2010, and focusing on 3 weak but intriguing candidate GRB neutrinos
(see Ref.\cite{icecubeCERN,icecubetesi}):
a 1.3 TeV neutrino  1.95$^o$ off GRB090417B with detection time  2249 seconds before the  trigger of GRB090417B,
a 3.3 TeV neutrino  6.11$^o$ off GRB090219 and detection time  3594 seconds before the GRB090219 trigger,
and a 109 TeV neutrino 0.2$^o$  off GRB091230A and detection time some 14 hours before the GRB091230A trigger.
The analysis reported in  Ref.\cite{gacGuettaPiran} would have been more intriguing
if the 109 TeV event could be viewed as a promising cosmological-neutrino candidate, but for that event
there was a IceTop-tank trigger coincidence. A single IceTop-tank trigger
is not enough to firmly conclude that the event was part of a cosmic-ray air shower, but of course that casts a shadow
 on the interpretation of the 109-TeV event as a GRB neutrino.

Unaware of the observations reported in Ref.\cite{gacGuettaPiran}, recently Stecker {\it et al.} reported in Ref.\cite{steckerliberati}
an observation which also might encourage speculations about neutrino propagation in quantum spacetime.
Ref.\cite{steckerliberati} noticed that IceCube data are presently consistent with a $\sim 2 PeV$ cutoff
for the cosmological-neutrino spectrum, and that this could be due
to novel processes (like ``neutrino splitting"\cite{steckerliberati,gacLRR})
that become kinematically allowed in the same class of quantum-spacetime models considered in  Ref.\cite{gacGuettaPiran}.

The study we are here reporting was motivated by these previous observations of Refs.\cite{gacGuettaPiran}
and \cite{steckerliberati}. Like Ref.\cite{gacGuettaPiran} our focus is on the hypothesis of GRB neutrinos with quantum-spacetime properties,
also exploiting the fact that, while Ref.\cite{gacGuettaPiran} was limited to IceCube data up to May 2010, the amount of data
now available from  IceCube \cite{IceCube} is significantly larger. Conceptually the main issue we wanted to face is indeed related
to the amount of IceCube data: as studies like these start to contemplate larger and larger groups of ``GRB-neutrino candidates"
some suitable techniques of statistical analysis must be adopted, and (unlike Refs.\cite{gacGuettaPiran}
and \cite{steckerliberati}) we wanted to devise a strategy of analysis applicable not only to one ``preferred model", but to a rather
 wide class of scenarios for the properties of the laws of propagation of neutrinos in a quantum spacetime.

As discussed more quantitatively below, the effects on propagation due to spacetime quantization can
be systematic or of ``fuzzy" type. Combinations of systematic effects and fuzziness are
also possible, and this is the hypothesis most challenging from the viewpoint of data analysis.
We came to notice that
 in all these scenarios    one should anyway find a correlation between the energy of the observed GRB neutrino and the
    difference between the time of observation of that neutrino and the trigger time of the relevant GRB.
Intriguingly our data analysis finds a rather strong such correlation, and we therefore argue that our findings should motivate a vigorous program of investigation following the strategy here advocated.

\section{Quantum-spacetime-propagation models and strategy of analysis}
The class of scenarios we intend to contemplate finds motivation in
some much-studied models of
spacetime quantization (see, {\it e.g.}, \cite{jacobpiran,gacsmolin,gacLRR,grbgac,gampul,urrutia,gacmaj,myePRL} and references therein)
 and, for the type of data analyses we are interested in, has the implication that
 the time needed for a ultrarelativistic particle\footnote{Of course the only regime of particle propagation
 that is relevant for this manuscript is the ultrarelativistic regime, since photons have no mass and
 for the neutrinos we are contemplating (energy of
 tens or hundreds of TeVs) the mass is completely negligible.}
to travel from a given source to a given detector receives a quantum-spacetime correction, here denoted with $\Delta t$.
We focus on the class of scenarios whose predictions for energy ($E$) dependence of $\Delta t$ can all be described 
in terms of the formula
(working in units with the speed-of-light scale ``$c$" set to 1)
\begin{equation}
\Delta t = \eta_X \frac{E}{M_{P}} D(z) \pm \delta_X \frac{E}{M_{P}} D(z) \, .
\label{main}
\end{equation}
Here the redshift- ($z$-)dependent  $D(z)$ carries the information on the distance between source and detector, and it factors
in the interplay between quantum-spacetime effects and the curvature of spacetime.
As usually done in the relevant literature \cite{jacobpiran,gacsmolin,gacLRR} we take for $D(z)$ the following form:\footnote{The interplay between quantum-spacetime effects and curvature of spacetime is still a lively subject of investigation, and, while (\ref{dz})
is by far the most studied scenario, some alternatives to (\ref{dz}) are also under consideration \cite{dsrfrw}.}
\begin{equation}
D(z) = \int_0^z d\zeta \frac{(1+\zeta)}{H_0\sqrt{\Omega_\Lambda + (1+\zeta)^3 \Omega_m}}  \, ,
\label{dz}
\end{equation}
where $\Omega_\Lambda$, $H_0$ and $\Omega_0$ denote, as usual,
respectively the cosmological constant, the Hubble parameter and the matter fraction, for which we take the values given in Ref.\cite{PlanckCosmPar}.
With $M_{P}$ we denote the Planck scale ($\simeq 1.2\,\cdotp 10^{28}eV$) while
the values of the  parameters $\eta_X$ and $\delta_X$ in (\ref{main})
characterize the specific scenario one intends to study. In particular, in (\ref{main}) we used the notation ``$\pm \delta_X$"
to reflect the fact that $\delta_X$ parametrizes the size of quantum-uncertainty (fuzziness) effects. Instead the parameter $\eta_X$
characterizes systematic effects: for example in our conventions for positive $\eta_X$ and $\delta_X =0$ a high-energy neutrino
is detected systematically after a low-energy neutrino (if the two neutrinos are emitted simultaneously).

The dimensionless parameters $\eta_X$ and $\delta_X$ can take different values for different particles \cite{gacLRR,myePRL,mattiLRR,szabo1}, and it is of particular interest for our study that in particular for
neutrinos some arguments have led to the expectation of an helicity dependence of the effects (see, {\it e.g.},
Refs.\cite{gacLRR,mattiLRR} and references therein). Therefore even when focusing only on neutrinos one should
contemplate four parameters, $\eta_+$, $\delta_+$, $\eta_-$, $\delta_-$ (with the indices $+$ and $-$ referring of course
to the helicity).
The parameters $\eta_X,\delta_X$ are to be determined experimentally. When non-vanishing,
they are expected to take values somewhere in a neighborhood of 1, but values as large as $10^3$ are plausible if the solution
to the quantum-gravity problem is somehow connected with the unification of non-gravitational forces \cite{gacLRR,wilczek,hsuHIGGSES}
while values
 smaller than 1 find support in some renormalization-group arguments (see, {\it e.g.}, Ref.\cite{hsuHIGGSES2}).

Presently for photons the limits on $\eta_\gamma$ and $\delta_\gamma$ are at the level of $|\eta_\gamma| \lesssim 1$
and $\delta_\gamma \lesssim 1$ \cite{fermiNATURE,gacNATUREPHYSICS2015}, but for neutrinos we are still several orders of magnitude below 1 \cite{steckerliberati,gacLRR}.
This is mainly
due to the fact that the observation of cosmological neutrinos is rather recent, still without any firm
identification of a source of cosmological neutrinos, and therefore the limits are obtained from terrestrial
experiments\footnote{Supernova 1987a was rather close by astrophysics standards and the signal detected in neutrinos
was of relatively low energy.}
(where the distances travelled are of course much smaller than the ones relevant in astrophysics).

For reasons that shall soon be clear we find convenient to introduce a ``distance-rescaled time delay" $\Delta t^*$ defined as
\begin{equation}
\Delta t^* \equiv \Delta t \frac{D(1)}{D(z)}
\label{tstar}
\end{equation}
so that (\ref{main}) can be rewritten as
\begin{equation}
\Delta t^* = \eta_X \frac{E}{M_{P}} D(1) \pm \delta_X \frac{E}{M_{P}} D(1) \, .
\label{maintwo}
\end{equation}
This reformulation of (\ref{main}) allows to describe the relevant quantum-spacetime effects,
which in general depend both on redshift and energy, as effects that depend exclusively on energy,
through the simple expedient of focusing on the relationship between $\Delta t$ and energy
when the redshift has a certain chosen value, which in particular we chose to be $z=1$.
If one measures a certain $\Delta t$ for a candidate GRB neutrino and the redshift $z$ of
the relevant GRB is well known, then one gets a firm determination of $\Delta t^*$ by simply rescaling
the measured $\Delta t$ by the factor $D(1)/D(z)$. And even when the redshift
of the relevant GRB is not known accurately one will be able to convert a measured $\Delta t$
into a determined $\Delta t^*$ with accuracy governed by how much one is able to still assume
about the redshift of the relevant GRB. In particular, even just the information on whether a GRB is long
or short can be converted into at least a very rough estimate of redshift.

Of course a crucial role is played in analyses such as ours by the criteria for selecting GRB-neutrino candidates.
We need a temporal window (how large can the $\Delta t$ be in order for us to consider a
IceCube event as a potential GRB-neutrino candidate) and we need criteria of directional selection (how well the directions estimated
for the IceCube event and for the GRB should agree in order for us to consider that IceCube event as a potential GRB-neutrino candidate).
While our analysis shall not include the above-mentioned 109-TeV neutrino (from Ref.\cite{gacGuettaPiran}), we do use it
to inspire a choice of the temporal window: assuming a 109-TeV GRB neutrino could be detected within 14 hours of the relevant GRB
trigger time, an analysis involving neutrinos with energies up to 500 TeV should allow for a temporal window of about 3 days,
and an analysis involving neutrinos with energies up to, say, 1000 TeV should allow for a temporal window of about 6 days.
Considering the rate of GRB observations of about 1 per day, we opt for focusing on neutrinos with energies
between 60 TeV\footnote{The 60-TeV lower limit of our range of energies is consistent with the analogous choice made by other studies
whose scopes, like ours, require keeping the contribution of background neutrinos relatively low \cite{IceCube,IceCubeBackground}.} and 500 TeV, allowing
for a temporal window of 3 days. Widening the range of energies up to, say, 1000 TeV would impose us indeed a temporal window
of about 6 days, rendering even more severe one of the key challenges for this sort of analysis, which is the one of multiple
GRB candidates for a single IceCube event. As directional criteria for the selection of GRB-neutrino candidates we consider the signal direction PDF depending on the space angle difference between GRB and neutrino: $P(\nu,GRB)=(2\pi\sigma^2)^{-1}\exp(-\frac{|\vec{x}_{\nu}-\vec{x}_{GRB}|^2}{2\sigma^2})$, a two dimensional circular Gaussian whose standard deviation is  $\sigma=\sqrt{\sigma_{GRB}^2+\sigma_{\nu}^2}$, asking
 the pair composed by the neutrino and the GRB to be at angular distance compatible within a 2$\sigma$ region.

A key observation for our analysis is that whenever $\eta_+$, $\eta_-$, $\delta_+$, $\delta_-$ do not vanish one should expect
on the basis of (\ref{maintwo}) a correlation between the $|\Delta t^*|$ and the energy of the candidate GRB neutrinos.
The interested reader will immediately see that this is obvious when $\delta_+ =\delta_- = 0$. It takes only
a little bit more thinking to notice that such a correlation should be present also when $\delta_+ \neq 0$ and/or $\delta_- \neq 0$
with $\eta_+ =\eta_- = 0$, as a result of how the fuzzy effects have range
that grows with the energy of the GRB neutrinos. We provide support for this conclusion in Fig.1.

\begin{figure}[t]
\includegraphics[scale=0.4]{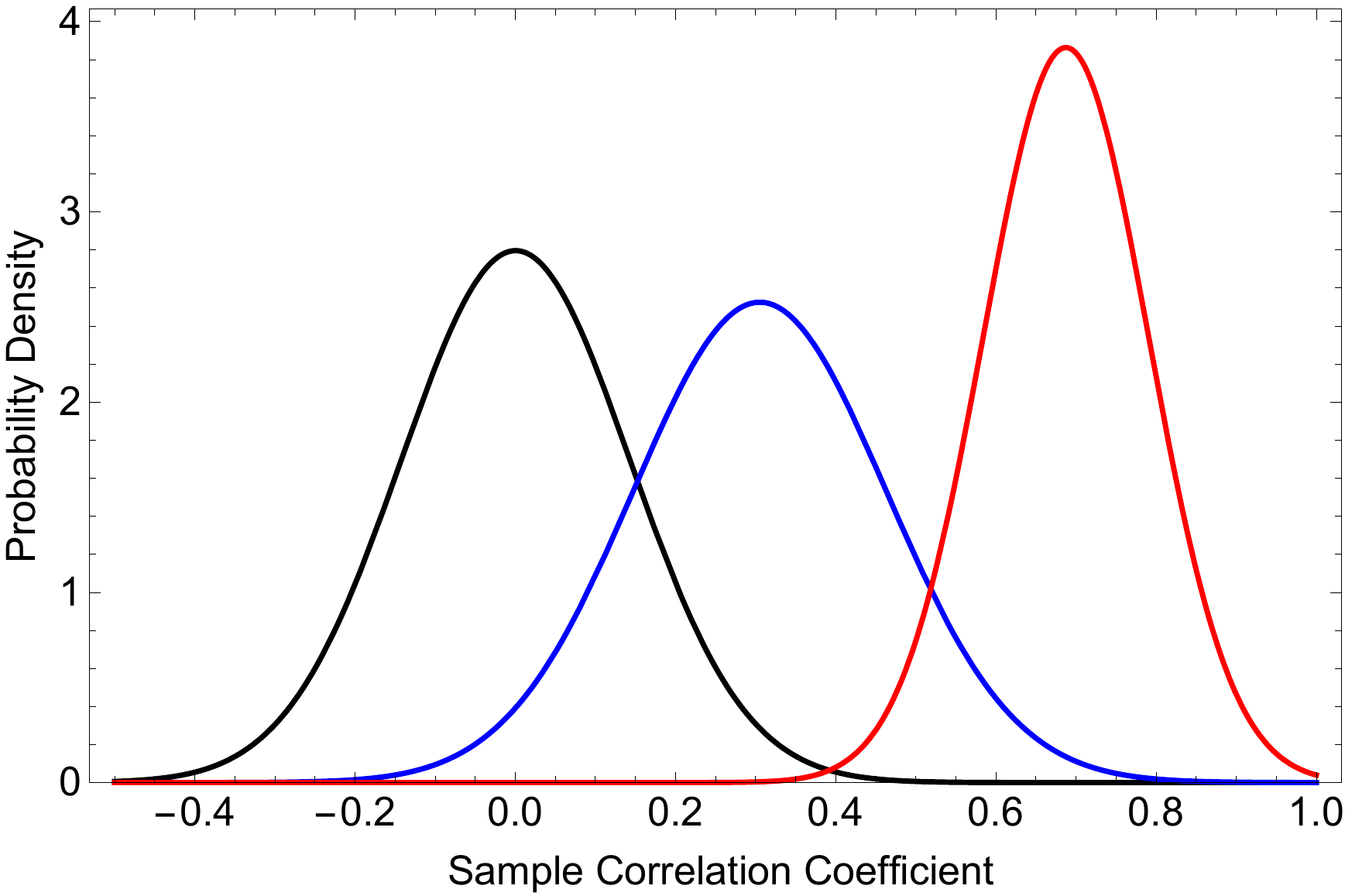}
\caption{Here we illustrate the different expectations one should have for the correlation on which we focus,
assuming all neutrinos are just background neutrinos (black), assuming $10\%$ of neutrinos are background while $90\%$ are GRB
neutrinos with $\eta_+ =\eta_- =0$, $\delta_+ =\delta_- =5$ (blue), or assuming $10\%$ of neutrinos are background while $90\%$ are GRB
neutrinos with $\eta_+ =\eta_-  =15$, $\delta_+ =\delta_- =5$ (red).
The probability densities were computed assuming
that the spectrum of the neutrinos decreases quadratically with energy ($E^{-2}$) between 60 and 500 TeV,
that the neutrinos would be observed only if within a 3-day window of the relevant GRB,
and, for simplicity, that all relevant GRBs are exactly at redshift of 1. This probability densities were obtained for the
hypothetical case of 50 candidate GRB neutrinos. The figure shows that 50 candidate GRB neutrinos would be enough
for the most likely correlation
outcome in the scenario with  $\eta_+ =\eta_- =0$, $\delta_+ =\delta_- =5$ to be
a rather unlikely outcome for the ``pure-background hypothesis." Actually,
much less than 50 candidate GRB neutrinos would be enough for the most likely correlation
outcome in the scenario with $\eta_+ =\eta_-  =15$, $\delta_+ =\delta_- =5$
to be a very unlikely outcome for the pure-background hypothesis.}
\end{figure}

\newpage

$\,$

\newpage

\section{RESULTS}

\vspace{-0.1cm}

Our data set\footnote{Both IceCube-neutrino data and GRB data used for this study were gathered from https://icecube.wisc.edu/science/tools} is for four years of operation of IceCube \cite{IceCube},
  from June 2010 to May 2014.
Since the determination of the energy of the neutrino plays such a  crucial role in our analysis
we include only IceCube ``shower events" (for ``track events" the reconstruction of the neutrino energy is far more problematic  and less reliable \cite{TRACKnogood1}).
We have 21 such events within our 60-500 TeV energy window, and we find that 9 of them
fit the requirements introduced in the previous section for candidate GRB neutrinos.
The properties of these 9 candidates that are most relevant for our analysis are summarized in Table 1 and Figure 2.

\vspace{-0.2cm}

\begin{table}[htbp]
\centering
{\def\arraystretch{0.3}\tabcolsep=3pt
\begin{tabular}{c|c|c|l|r|c}
\hline
$\,$                   &    \!\!\!   E \!\!\!\! [TeV]      \!\!\!        & GRB              & z           & $\Delta t^*$ [s]      & $\,$   \\\hline \hline
IC9                          &                 63.2                 & 110503A    & 1.613        & 50227       &     *     \\\hline
IC19                         &                71.5                  & 111229A    & 1.3805        & 53512       &     *   \\\hline
\multirow{3}{*}{IC42} & \multirow{3}{*}{76.3}    & 131117A      & 4.042    & 5620       &            \\
&                                       & 131118A      & 1.497  *  &  -98694       &     *      \\
&                                       & 131119A      & \,\,\,\,\, ? &  -146475     &             \\
\hline
IC11                          &                88.4                 & 110531A    & 1.497   *     & 124338       &     *     \\\hline
IC12                         &                104.1                 & 110625B    & 1.497   *     & 108061       &     *    \\\hline
\multirow{3}{*}{IC2} & \multirow{3}{*}{117.0}    & 100604A      & \,\,\,\,\, ?  & 10372       &            \\
                                &                                       & 100605A      & 1.497 *    &  -75921       &     * \\
                                &                                       & 100606A      & \,\,\,\,\, ?  &  -135456     &                   \\\hline

IC40                          &             157.3                 & 130730A    & 1.497   *     & -120641       &     *    \\\hline
\multirow{2}{*}{IC26}
& \multirow{2}{*}{210.0}   & 120219A      & 1.497  *  &  153815     &     *       \\
&                          & 120224B     & \,\,\,\,\, ? &  -117619    &           \\
\hline
IC33                          &             384.7                 & 121023A    & \,\, 0.6      *     &  -289371       &     *     \\\hline
\end{tabular}
}
\caption{Among the 21 ``shower neutrinos" with energy between 60 and 500 TeV observed by IceCube between June 2010 and May 2014
only 9 fit our directional and temporal criteria for GRB-neutrino candidates, and yet for 3 of them there is more than one GRB
to be considered when pairing up neutrinos and GRBs. The last column highlights with an asterisk the 9 GRB-neutrino candidates ultimately selected
by our additional criterion of maximal correlation. Also shown in table are the values of redshift attributed to the relevant GRBs:
the redshift is known only
for GRB111229A and GRB110503A (plus GRB131117A, which however ends up not being among the GRBs selected by the maximal-correlation criterion).
GRB111229A and GRB110503A are long GRBs and we assume that the average of their redshifts (1.497) could be a reasonably good estimate
of the redshifts of the other long GRBs relevant for our 9 GRB-neutrino candidates. These are the 6 estimated values of redshift
$z=1.497^*$, the asterisk reminding that it is a ``best guess" value. For analogous reasons we place an asterisk close to the value of 0.6 which is our best guess for the redshift of the only short GRB
in our sample. The first column lists the ``names" given by IceCube to the neutrinos that end up being relevant for our analysis.
Also notice that 5 of our GRB-neutrino candidates are ``late neutrinos" ($\Delta t^*>0$),
while the other 4 are ``early neutrinos" ($\Delta t^*<0$): this might be of interest to some readers but plays no role in our study
since our correlations involve the absolute value of $\Delta t^*$.}
\label{table1}
\end{table}

\begin{figure}[h!]
\includegraphics[scale=0.3]{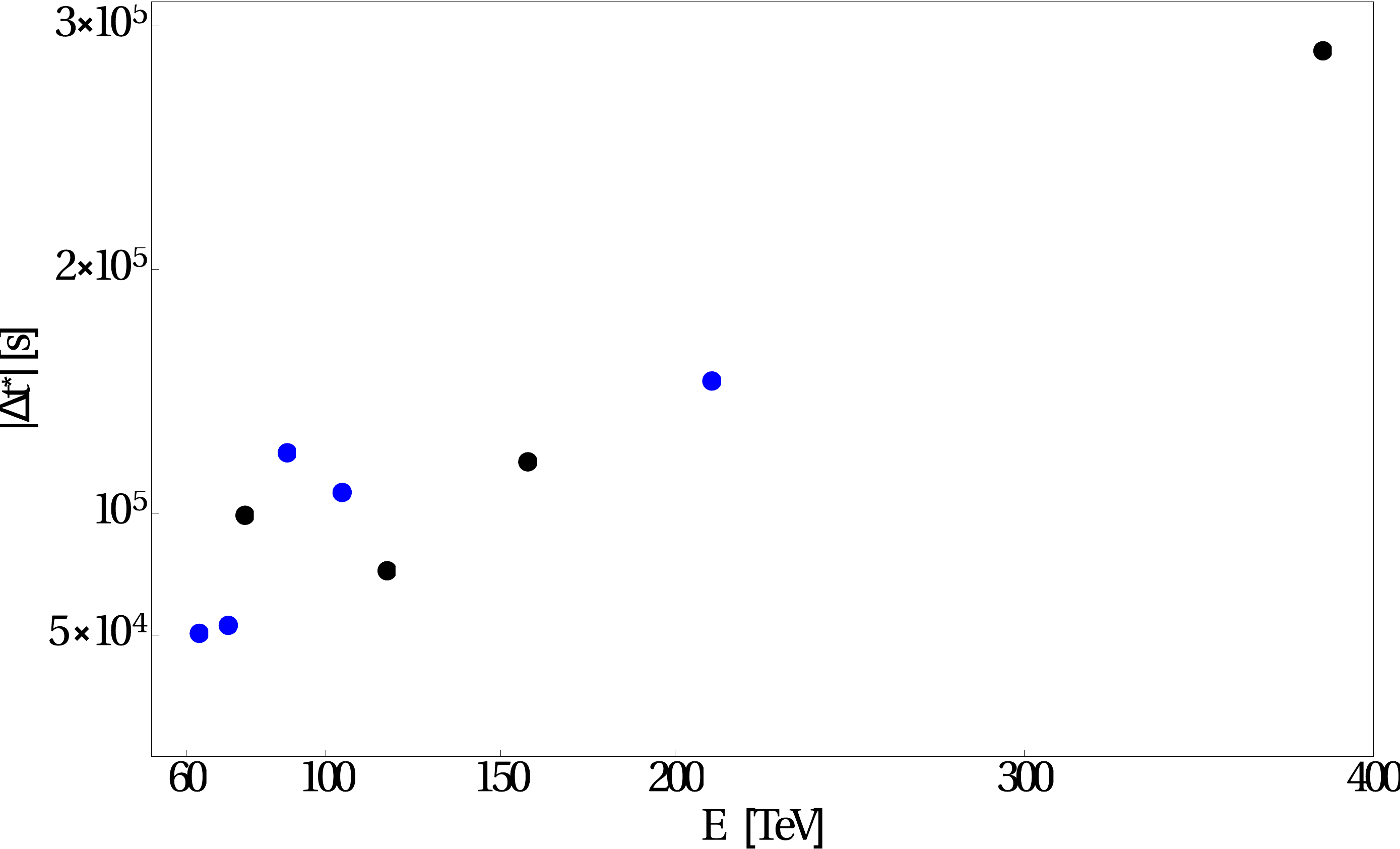}
\caption{Points here in figure correspond to the 9 GRB-neutrino candidates highlighted with an asterisk in the last column of Table 1. Blue points correspond to ``late neutrinos" ($\Delta t^*>0$),
while black points correspond to ``early neutrinos" ($\Delta t^*<0$).}
\end{figure}

In commenting Table 1
we start by noticing that
for some IceCube events our selection
criteria produce multiple GRB-neutrino candidates (and the situation would have been much worse if we had considered a wider energy range
and a correspondingly wider temporal window).
Since we have two cases with 3 possible GRB partners and one case with a pair of possible GRB partners, we must contemplate 18
alternative descriptions of our 9 GRB-neutrino candidates.
As neutrino telescopes gradually accrue more and more such events the number of combinations to be considered
in analyses such as ours will grow very large.
We propose that in general this issue of multiple candidates should be handled, consistently with the nature of the hypothesis being tested,
by focusing on the case
 that provides the highest correlation. This might appear to introduce a bias toward higher values of the correlation,
but, as we shall soon argue, the significance of such an analysis is not given by the correlation itself but rather requires the evaluation
of a ``false alarm probability", and for the false alarm probability this criterion for handling multiple candidates introduces
no bias (see below).

Another issue reflected by Table 1 comes from the fact that for only 3 of the GRBs involved in this analysis the redshift is known.  We must handle only one short GRB of unknown redshift, and we assume for it a redshift of 0.6,
which is a rather reasonable rough estimate for a short GRB
(but we shall contemplate also values of 0.5 and of 0.7).
For some of our long GRBs we do have a redshift determination and we believe that consistently with the hypothesis here being tested
one should use those known values of redshift for obtaining at least a rough estimate of the redshift of long GRBs for which
the redshift is unknown. This is illustrated by the 9 GRB-neutrino candidates marked by an asterisk in table 1:
those 9 candidates include 8 long GRBs, 2 of which have known redshift, and we assign to the other 6 long GRBs
the average ${\bar{z}}$  of those two values of redshift (${\bar{z}}=1.497$). As it will be reported in the PhD thesis of Ref.\cite{Juniorthesis}, we have checked that our results do not depend
strongly on the what is assumed about unknown redshifts, be it assuming that these redshifts follow the distribution of GRBs observed
 in photons or simply assuming different values of ${\bar{z}}$. We shall document a bit of this insight here below,
 by providing our results both assuming this criterion of the ${\bar{z}}$ and assuming simply a redshift of 2 for all
 long GRBs of unknown redshift.
We feel that estimating a ${\bar{z}}$  from the ``data points"
is the only reasonable way to proceed, since we do not expect that the redshift distribution of GRBs observed also in neutrinos
should look much like the redshift distribution of GRBs observed only in photons. However we imagine
that some readers might have been
more comfortable if we assumed for our long GRBs of unknown redshift the average value of redshift
of GRBs observed in photons, which is indeed of about 2.

Having specified these further prescriptions, we can proceed to compute
the correlation between $|\Delta t^*|$ and energy
 for our 9 GRB-neutrino candidates.
Because of the fact that for some of our neutrinos there is more than one possible GRB partner we end up having 18 such values
of correlation, and remarkably they are all very high: the highest of these 18 values
is of 0.951 (the corresponding 9 neutrino-GRB pairs are highlighted by an asterisk in Table 1 and are shown in Figure 2),
and even the lowest of these 18 values of correlation is still of 0.802.
  In Table 2 we show how the evaluation of the maximum correlation for our 9 GRB-neutrino candidates
  would change
 upon replacing our ${\bar{z}}$ with a redshift of $2$, for long GRBs, and upon replacing the  value of 0.6
 we assumed for the redshift of the short GRB in our collection with 0.5 or 0.7.

\vspace{-0.2cm}
\begin{table}[htbp]
\center
\begin{tabular}{c|c|c|c}
%\diagbox[width=1cm]
$ \, $	                          & $z_{long} =\bar{z}$ 	& $z_{long} = 2$	\\ \hline
${z_{short}}=0.5$		    &  0.958 	                      & 0.953 	\\ \hline
${z_{short}}=0.6$			&  {\bf 0.951} 	              & 0.960 	\\ \hline
${z_{short}}=0.7$			&  0.941 	              & 0.964 	\\
\end{tabular}
\label{table2}
\caption{Adopting our  ``$\bar{z}$ criterion" for long GRBs whose redshift is not known and z=0.6 for short GRBs one gets as maximal correlation for our data the impressive value of 0.951. Here we show how this estimate changes if one assigns to short GRBs
the alternative values of redshift of 0.5 and 0.7 and/or one replaces our $\bar{z}$ with a redshift of 2.}
\end{table}
The class
of quantum-spacetime scenarios we are considering predicts a non-vanishing (and possibly large) correlation,
and we did find on data very high values of correlation.
This in itself however does not quantify what is evidently the most interesting quantity here of interest,
which must be some sort of ``false alarm probability": how likely it would be to have accidentally data with such
good agreement with the expectations of the quantum-spacetime models here contemplated?
We need to estimate how often
a sample composed exclusively of background neutrinos\footnote{Consistently with the objectives of our analysis we consider as ``background neutrinos" all
 neutrinos that are unrelated to a GRB, neutrinos of atmospheric or other astrophysical origin which end up being selected
as GRB-neutrino candidates just because accidentally their time of detection and angular direction happen to fit
our selection criteria.} would produce accidentally
9 or more GRB-neutrino candidates with correlation
comparable to (or greater than) those we found in data.
We do this by performing $10^5$ randomizations of the times of detection
of the 21 IceCube neutrinos relevant for our analysis, keeping their energies fixed,
and for each of these time randomizations we redo the analysis just as if they were real data.
Our observable is a time-energy correlation and by randomizing the
times we get a robust estimate of how easy (or how hard) it is for a sample composed exclusively
of background neutrinos to produce accidentally
a certain correlation result.
In the analysis of these fictitious data obtained by randomizing the detection times of the neutrinos
we handle cases with neutrinos for which there is more than one possible GRB partner by maximizing the correlation,
in the sense already discussed above for the true data. We ask how often this time-randomization
procedure produces 9 or more GRB-neutrino candidates with correlation
$\geq 0.951$, and remarkably we find that this happens only in 0.03$\%$ of cases.

In Table 3 we report a preliminary investigation of
how this result of a 0.03$\%$ false alarm probability depends
on the assumptions we made for redshifts.
 Table 3 is in the same spirit of what was reported in our Table 2 for the estimates of the correlation.
 Each entry in Table 2 recalculates the false alarm probability just like we did above to obtain the
 result of  0.03$\%$, but now considering some alternative possibilities for the assignment of redshifts
 to GRBs whose redshift is actually unknown.
Once again for long GRBs we consider two possibilities, the ${\bar{z}}$ discussed above and redshift of $2$,
 while for short GRBs we consider values of redshift of 0.5, 0.6
 and 0.7. Table 3 shows that our false alarm probability does not change much within this range of exploration
 of the redshift assignments.

\begin{table}[htbp]
\center
\begin{tabular}{c|c|c|c}
%\diagbox[width=1cm]
$ \, $	& $z_{long} =\bar{z}$ 	& $z_{long} = 2$	\\ \hline
${z_{short}}=0.5$			&  0.03 \% 	& 0.04 \%	\\ \hline
${z_{short}}=0.6$			&  {\bf 0.03 \%}	& 0.02  \%	\\ \hline
${z_{short}}=0.7$			&  0.04 \%  & 0.01 	\% \\
\end{tabular}
\label{table3}
\caption{Adopting our  ``$\bar{z}$ criterion" for long GRBs whose redshift is not known and z=0.6 for short GRBs one gets a
false alarm probability of 0.03$\%$. Here we show how this estimate changes if one assigns to short GRBs
the alternative values of redshift of 0.5 and 0.7 and/or one replaces our $\bar{z}$ with a redshift of 2.}
\end{table}

Our next objective is to see how things change if one is ``unreasonably conservative"
in assessing the implications of our prescription for handling
cases where there is more than one possible GRB partner for a neutrino.
We are proposing that one should address this multi-candidate issue in the way
that maximizes the correlation, and this evidently introduces some bias toward
higher values of the correlation. However, as already stressed above,
when we randomize (fictitious)
detection times we handle the multi-candidate issue in exactly the same way, by maximizing the correlation,
so that overall there is no bias
for the false alarm probability.
It is nonetheless interesting to notice that
one still obtains a rather low false alarm probability
even when comparing the minimum correlation for our true data to the maximum correlation
for the fictitious data obtained by randomizing neutrino detection times.
So we now ask how often the fictitious data obtained by randomizing neutrino detection times
produce 9 or more GRB-neutrino candidates with correlation $\geq 0.803$ ($0.803$ being, as noticed above, the lowest possible
 value of correlation for our true data), but for the fictitious data we still
 handle cases with neutrinos having more than one possible GRB partner
 by maximizing the correlation.
 Even this procedure, which is evidently biased toward lower values of the false alarm probability,
 only gives a false alarm probability of $\simeq 1\%$.
Table 4 explores the dependence on assumptions for redshift of the value of $0.803$ for the lowest correlation obtainable from the true data,
while Table 5 explores analogously the dependence on assumptions for redshift of our result for the
``unreasonably conservative estimate of the false alarm probability."

\begin{table}[htbp]
\center
\begin{tabular}{c|c|c|c}
%\diagbox[width=1cm]
$ \, $	                          & $z_{long} =\bar{z}$ 	& $z_{long} = 2$	\\ \hline
${z_{short}}=0.5$		    &  0.844 	                      & 0.869 	\\ \hline
${z_{short}}=0.6$			&  {\bf 0.803} 	              & 0.849 	\\ \hline
${z_{short}}=0.7$			&  0.751                     & 0.822 	\\
\end{tabular}
\label{table2}
\caption{Adopting our  ``$\bar{z}$ criterion" for long GRBs whose redshift is not known and z=0.6 for short GRBs one gets as minimal correlation for our data a still high value of 0.803. Here we show how this estimate changes if one assigns to short GRBs
the alternative values of redshift of 0.5 and 0.7 and/or one replaces our $\bar{z}$ with a redshift of 2.}
\end{table}

\vspace{-0.2cm}
\begin{table}[htbp]
\center
\begin{tabular}{c|c|c|c}
%\diagbox[width=1cm]
$ \, $	& $z_{long} =\bar{z}$ 	& $z_{long} = 2$	\\ \hline
${z_{short}}=0.5$			&  0.7 \% 	& 0.6 \%	\\ \hline
${z_{short}}=0.6$			&  {\bf 1.0 \%}	   & 0.6  \%	\\ \hline
${z_{short}}=0.7$			&  1.5 \%  & 0.8	\% \\
\end{tabular}
\label{table4}
\caption{Adopting our  ``$\bar{z}$ criterion" for long GRBs whose redshift is not known and z=0.6 for short GRBs we obtain
an ``unreasonably conservative estimate of the false alarm probability" which is still only 1.0$\%$.
 Here we show how this estimate changes if one assigns to short GRBs
the alternative values of redshift of 0.5 and 0.7 and/or one replaces our $\bar{z}$ with a redshift of 2.}
\end{table}

\section{Toward estimating model parameters}
In searching for evidence of quantum-spacetime effects on neutrino propagation our approach has the advantage
of allowing to study at once a variety of scenarios, the scenarios obtainable by all sorts of combinations
of values for  $\eta_+ $, $\eta_-$, $\delta_+$, $\delta_-$.
This is due to the fact that positive correlation between $\Delta t^*$ and $E$ is expected whenever
one or more of the parameters  $\eta_+ $, $\eta_-$, $\delta_+$, $\delta_-$ are non-zero.
Our approach performs very well in comparing the hypothesis ``all the GRB-neutrino candidates actually are
background neutrinos" to the hypothesis ``some of the GRB-neutrino candidates truly are GRB neutrinos
governed by Eq.(1) with one or more of the parameters $\eta_+ $, $\eta_-$, $\delta_+$, $\delta_-$ having non-zero value."
It does so in ways that are rather robust with respect to the assumptions made about the redshift
of the relevant GRBs and with respect to the presence of some background neutrinos among the GRB-neutrino candidates.

Our false-alarm probabilities are still not small enough to worry about that,
but if it happens that the experimental situation develops positively for our scenario then
one will of course be interested in estimating model parameters, {\it i.e.}
comparing how well different choices of values of the parameters of the model match the available data.
This is clearly harder within our approach. In particular it surely requires some reasonable 
estimate of the amount of background neutrinos
present among the GRB-neutrino candidates.
Testing the hypothesis that all the GRB-neutrino candidates actually are
background neutrinos is evidently simpler than testing the hypothesis
that some of the GRB-neutrino candidates are background and some other
are truly GRB neutrinos: for the latter one would need to specify  how many
are background and how many are GRB neutrinos.

While we postpone contemplating these issues until (and if) the experimental situation
evolves accordingly, we still find appropriate to offer here at least
a rudimentary attempt of interpretation of the data on the basis of the
parameters of our reference Eq.(1), assuming naively that all our GRB-neutrino candidates 
actually are GRB neutrinos.
Because of the accordingly explorative nature of the observations reported in this section, we shall be satisfied taking
as reference the 9 GRB-neutrino candidates marked with an asterisk in Table 1 and considered in Fig.2, {\it i.e.}
the maximum-correlation 9-plet. If the experimental situation develops in such a way to provide motivation for
more refined estimations of model parameters, the relevant procedures should not only rely on some estimate of the amount
of background neutrinos but should also handle the fact that some neutrinos have more than one
 possible GRB partner, in the same spirit we adopted for the estimates of the false-alarm probability given in the previous
section. At the present stage we find sufficient not only to neglect background neutrinos and consider exclusively
the maximum-correlation 9-plet, but also to focus on a simplified version of the phenomenological model.
As first simplification we assume $\eta_+ + \eta_- =0$, which is 
reasonably consistent with the fact that in Fig.2 one sees about an equal number of
candidate ``early neutrinos" and candidate ``late neutrinos".
In addition we further restrict our attention to the case $\delta_+ = \delta_-$, so that we must only be concerned
 with the parameters $\eta_+$ and $\delta_+$ (with then $\eta_- =- \eta_+$, $\delta_- =\delta_+$).

 Having specified these restrictions we first take a very simple-minded approach and assume that
 the features shown in Fig.2 are all due to Eq.(1). This in particular
 means we are naively assuming that there are no background neutrinos, that the estimates of GRB redshifts given in table 1 are exact,
 and that points in Fig.2 fail to be on a straight line exclusively because of the effects of the parameter $\delta_+$
 (and $\delta_-$, with $\delta_-=\delta_+$).
 This leads to  $|\eta_+| = 22 \pm 2 $ and $\delta_+ = 6 \pm 2$.
%  (taking into account that the estimate $|\eta_+| = 22 \pm 2$ was obtained from 9
% data points).

Next we perform a Bayesian analysis to derive posterior distributions of unknown parameters. 
We assume again simple-mindedly that there are no background neutrinos, and
we handle as unknown parameters
not only the parameters of our model, $|\eta_+|$ and $ \delta_+$, but also the standard deviation $\delta z$  of the normal distribution
 that we tentatively assume to describe the redshift distribution of long GRBs observed also in neutrinos. As mean value
 of this normal distribution we take 1.497, following the argument discussed in the previous section.
 For the redshift distribution of short GRBs observed also in neutrinos (which is relevant for only one of our GRB-neutrino
  candidates) we simply assume a normal distribution with mean value 0.6 and standard deviation of 0.2. In order to evaluate the marginalized posterior probability density functions of the parameters $|\eta_+|$, $ \delta_+$ and $\delta z$ we use the Markov chain Monte Carlo technique, with uniform priors with ranges $0 \leq \eta \leq 50$, $0 \leq \delta \leq 10$ and $ 0 \leq \delta z \leq 1$. Uncertainties for the energies of the neutrinos (see Ref. \cite{IceCube}) were also taken into account.
This Bayesian analysis
determines  $\delta z$ to be $\delta z = 0.45 \pm 0.17$, and for the parameters of our model
gives $|\eta_+| = 23 \pm 2$, $\delta_+ = 4.7 \pm 1.5 $, which is consistent with what we had concluded in the previous paragraph ($|\eta_+| = 22 \pm 2 $, $\delta_+ = 6 \pm 2$) on the basis of more simple-minded considerations.

% So we conclude that, when restricting the class of models here considered to the case
% $\eta+ + \eta_- =0$, $\delta_+ = \delta_-$, the present experimental situation
% would lead to the estimates $|\eta+| = |\eta_-| = XXXXX \pm XXXXXX$
% and $\delta_+ = \delta_- = XXXXX \pm XXXXXX$. An analysis of broader scope, removing the
% simplifying assumptions $\eta+ + \eta_- =0$ and $\delta_+ = \delta_-$, will  be justified
% once more data are accrued by IceCube, if the feature here uncovered persists.

\section{OUTLOOK}
As mentioned, our work took off from the analogous study reported in Ref.\cite{gacGuettaPiran}, with additional motivation found in
what had been reported in Ref.\cite{steckerliberati}.
We looked within IceCube data from June 2010 to May 2014 for the same feature which had been already noticed
in Ref.\cite{gacGuettaPiran}, in an analysis based on much poorer IceCube data for the period from April 2008 to May 2010.
The study of Ref.\cite{gacGuettaPiran} was intriguing but ultimately
appeared to be little more than an exercise in data-analysis strategy, since it could
only consider 3 neutrinos, none of which could be viewed as a promising GRB-neutrino candidate. The 109-TeV event considered
in Ref.\cite{gacGuettaPiran} could be easily dismissed as likely the result of a cosmic-ray air shower, and the other
two neutrinos were of much lower energy, energies at which atmospheric neutrinos are very frequent. Yet what we found
here is remarkably consistent with what had been found in Ref.\cite{gacGuettaPiran}. Particularly the 109-TeV event
would be a perfect match for the content of our Figure 2, as the interested reader can easily verify.
We chose to rely exclusively on data unavailable to Ref.\cite{gacGuettaPiran},  IceCube data from June 2010 to May 2014,
 and on these
new data alone the feature is present very strongly, characterized by a false alarm probability 
which we estimated fairly at $0.03 \%$ and ultraconservatively at $1\%$.
We feel this should suffice to motivate a vigorous program of further investigation of the scenarios here analyzed.

Particularly over these last few decades of fundamental physics, results even more encouraging
than ours have then gradually faded away, as more data was accrued, and we are therefore well prepared to see our neutrinos
have that fate.
We are more confident that our strategy of analysis will withstand the test of time. The main ingredient of novelty
 is the central role played by the correlation between the energy of a neutrino
and the difference between the time of observation of that neutrino and the trigger time of a GRB.
The advantage of focusing on this correlation is that it is expected in a rather broad class of phenomenological models
of particle propagation in a quantum spacetime, which was here summarized in our Eq.(\ref{main}).
Moreover, by analyzing a few representative cases of simulated data we find \cite{Juniorthesis} that such correlation studies
are rather robust with respect to uncertainties in the estimates of the rates of background neutrinos,
and this could be valuable: extrapolating to higher energies known facts about the rate of atmospheric neutrinos is already
a challenge, but for analyses such as ours one would also need to know which percentage of cosmological
neutrinos are due to GRBs, an estimate which at present is simply impossible to do reliably.
Comparing for example our approach to
the strategy of analysis adopted in Ref.\cite{gacGuettaPiran}
one can see immediately that the strategy of analysis adopted in Ref.\cite{gacGuettaPiran}
is inapplicable when $\delta_X \neq 0$
(whether or not $\eta_X = 0$). When $\eta_X \neq 0$, $\delta_X =0$ we find \cite{Juniorthesis}
that our approach and the approach of Ref.\cite{gacGuettaPiran}
perform comparably well if the rate of background neutrinos is well known, but ours is indeed more
robust with respect to uncertainties in the estimates of the rates of background neutrinos.

\section*{Acknowledgements}

We are grateful to Antonio Capone for sharing with us some of his insight on neutrino telescopes
and we are grateful to Jerzy Kowalski-Glikman for some valuable comments on an earlier version
of this manuscript.
The work of GR  was supported  by funds provided by the National Science Center under the
agreement DEC- 2011/02/A/ST2/00294.
NL acknowledges support by the European Union Seventh Framework Programme (FP7 2007-2013) under grant agreement 291823 Marie Curie FP7-PEOPLE-2011-COFUND (The new International Fellowship Mobility Programme for Experienced Researchers in Croatia - NEWFELPRO).

\end{document}